# Digital image restoration based on pixel simultaneous detection probabilities

V. Grabski

*Instituto de Física Universidad Nacional Autonoma de Mexico, A.P. 20-364, 01000 DF, Mexico*

Here an image restoration on the basis of pixel simultaneous detection probabilities (PSDP) is proposed. These probabilities can be precisely determined by means of correlations measurement [NIMA 586 (2008) 314-326]. The proposed image restoration is based on the solution of matrix equation. Non-zero elements of Toeplitz block matrix with ones on the main diagonal, is determined using PSDP. The number of non zero descending diagonals depends on the detector construction and is not always smaller than 8. To solve the matrix equation, the Gaussian elimination algorithm is used. The proposed restoration algorithm is studied by means of the simulated images (with and without additive noise using PSDP for General Electric Senographe 2000D mammography device detector) and a small area (160x160 pixels) of real images acquired by the above mentioned device. The estimation errors of PSDP and the additive noise magnitude permits to restore images with the precision better than 3% for the above mentioned detector. The additive noise in the real image is present after restoration and almost has the same magnitude. In the restored small area (16x16 mm) of real images, the pixel responses are not correlated. The spatial resolution improvement is also analyzed by the image of an absorber edge.

*Index Terms*—Correlations, pixel simultaneous detection probabilities, point spread function, image restoration.

## I. Introduction

AN IMAGE acquired by the digital detector includes the degradation of the original image conditioned by the pixel simultaneous counting. The most frequently used technique for the restoration is based on the de-convolution using a two-dimensional point spread function [1,2] that is not easy to evaluate [3]. The restoration in the spatial frequency domain that requires MTF evaluation (on the base of PSF) is also not an easy task due to aliasing [4] and noise amplification in higher spatial frequencies [2]. In the pixel domain, the restoration requires the estimation of PSF and its integration in the neighboring pixel area, in order to perform the image restoration. This requires the exact knowledge of the pixel response function. These two steps can be combined if the above mentioned PSF integrals are estimated directly. In this case, the knowledge of pixel response function is not required. The possibility of the above mentioned estimation is already demonstrated in the study [5] where the integral ratios have been called as the pixel simultaneous detection probabilities (PSDP). The said study has also suggested the possibility of the usage of PSDP for the purpose of image restoration. This present work is a study of the image restoration precision that can be achieved, taking into account the estimation errors of PSDP and the presence of the additive noise. The main idea is to perform the restoration of the detector, blurring separately among the other sources of the image blurring. If successful, this will be an opportunity to use thick scintillation converters that are very important in digital radiography [6].

The detector blurring restoration process (without the use of minimization procedure) depends on three important factors: the first is the existence of the additive noise, the second is the estimation errors of PSDP and the third is the stability of the solution of matrix equation for a very large amount of pixel numbers.

The influence of the additive noise on the restoration process in pixel domain in radiography has been studied for a long time [2]. Here, the minimization procedure is used to find the closest approximation (or the most probable image) for the hidden image. In the detector blurring restoration, the quantum noise no longer has importance. Only the electronic (see section Method) noise and the fluctuations conditioned by the detection of optical photons produced in the converter are important. The latter is exposure dependent and can have significant variation depending on the image location. The influence of the additive noise on the precision of the restoration depends on the additive noise magnitude and the detector properties. Here all studies are performed by means of simulations using GE mammography device detector characteristics. The restoration procedure is also considered when the additive noise has an order of larger magnitude than the above mentioned detector electronic noise.

The influence of the precision of PSDP on the restoration process is not studied in the literature yet. Due to the enormous amount of operations during the restoration process, the small changes in these parameters probably can significantly affect the restored image. This study has also been performed by the simulations to check the restoration accuracy depending on the precision of PSDP estimation.

The stability of the matrix equation solution is connected with the loss of precision during the rounding process of the enormous amount of operations. There are several classes of algorithms for solving such systems: regular Gaussian elimination algorithms that exploit the Toeplitz matrix structure ($O(N^2)$ operations are required) and fast $O(N\log N)$ algorithms based on the usage of the fast Fourier transform [1]. The general theoretical limitations [7] are very rough and difficult to use for the estimation of the expected precisions. That's why in this work, the study of the stability of solutions and expected precision dependent on the accuracy of PSDP determination by means of simulation, has been performed.

For the validation of the proposed restoration algorithm, raw images for the beam energy 26-28 kV, have been acquired using GE Mammography unit 2000D device.

II. METHOD

If the initial photon number in pixel $(i,j)$ is $x_{ij}$ and the simultaneous detection probability for the same photon is $\alpha_{mn}$ (where $m = \pm 0, \pm 1,..$ and $n = \pm 0, \pm 1,..$ ), then the real value $y_{ij}$ detected in pixel $(i,j)$ (neglecting the additive noise contribution) can be written (accounting for image degradation) as:

$$y_{ij} = \sum_{m,n=-s}^{s} \alpha_{mn} x_{i+m \, j+n} , \qquad (1)$$

where $s$ is the maximum number of pixels around a given pixel $(i,j)$ when $\alpha_{mn} \neq 0$. Here we follow the considerations in the study [5] when 9-$\alpha$'s are non-zero for the detector GE Senographe 2000D device. Assuming that there is an inverse symmetry (see Fig 1), $\alpha_{mn}$ is listed in the Table [5].

Table Pixel simultaneous detection

probabilities

| $a_{00}$ | $a_{01}$ | $a_{10}$ | $a_{11}$ | $a_{-11}$ |
|---|---|---|---|---|
| 1 | 0.102±0.001 | 0.094±0.001 | 0.022±0.001 | 0.027±0.001 |

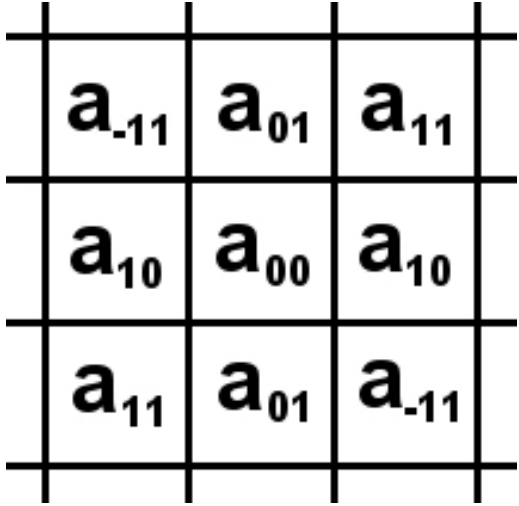

Fig 1 PSDP location on the matrix

This limitation is conditioned by the method of the estimation of $\alpha_{mn}$ [5] (though PSDP can be estimated without the above mentioned assumption acquiring many images in the same condition). The Eq (1) in matrix form can be represented as:

$$Y = AX, \quad (2)$$

where A is the Topeliz matrix NxN, Y and X are the vectors of size N and N is the number of the image points. The Matrix A has a block structure corresponding to the number of image columns and rows and with the ones on the main diagonal can be determined using $\alpha_{mn}$.

$$A = \begin{bmatrix} a_{00} & a_{10} & 0 & . & 0 & a_{01} & a_{-11} & 0 & . & . & . & . & . & 0 \\ a_{10} & a_{00} & a_{10} & 0 & 0 & a_{11} & a_{01} & a_{-11} & 0 & . & . & . & . & . \\ 0 & a_{10} & . & . & 0 & 0 & a_{11} & . & . & 0 & . & . & . & . \\ . & 0 & . & . & a_{10} & 0 & 0 & . & . & a_{-11} & 0 & . & . & . \\ 0 & 0 & 0 & a_{10} & a_{00} & 0 & 0 & 0 & a_{11} & a_{01} & 0 & 0 & . & . \\ a_{01} & a_{11} & 0 & 0 & 0 & a_{00} & a_{10} & 0 & 0 & 0 & a_{01} & a_{-11} & 0 & . \\ a_{-11} & a_{01} & a_{11} & 0 & 0 & a_{10} & a_{00} & a_{10} & 0 & 0 & a_{11} & . & . & 0 \\ 0 & a_{-11} & . & . & 0 & 0 & a_{10} & . & . & 0 & 0 & . & a_{-11} & 0 \\ . & 0 & . & . & a_{11} & 0 & 0 & . & . & a_{10} & 0 & 0 & a_{11} & a_{01} & a_{-11} \\ . & . & 0 & a_{-11} & a_{01} & 0 & 0 & 0 & a_{10} & a_{00} & 0 & 0 & 0 & a_{11} & a_{01} \\ . & . & . & 0 & 0 & a_{01} & a_{11} & 0 & 0 & 0 & a_{00} & a_{10} & 0 & 0 & 0 \\ . & . & . & . & 0 & a_{-11} & a_{01} & a_{11} & 0 & 0 & a_{10} & . & . & 0 & 0 \\ . & . & . & . & . & 0 & a_{-11} & . & . & 0 & 0 & . & . & a_{10} & 0 \\ . & . & . & . & . & . & 0 & . & a_{01} & a_{11} & 0 & 0 & a_{10} & a_{00} & a_{10} \\ 0 & . & . & . & . & . & . & 0 & a_{-11} & a_{01} & 0 & . & 0 & a_{10} & a_{00} \end{bmatrix}$$

The stability of solutions of Eq (2) depends on the size of the block structure ($\sim\sqrt{N}$) and the values of $a_{ij}$. In solving Eq (2), pivoting is not required and the multiplication coefficients are always smaller than 1. So the precision lost for the used algorithm depends on the block size and can be roughly estimated as a number of significant operations by $N^{1.5}$x(operation precision) which is still small even for images with pixel numbers of order $N\sim10^6$.

Considering the additive noise, the equation (2) is modified

$$Y = AX + G, \quad (3)$$

Where G is the vector of the additive noise and usually is unknown. A trial restoration $X^r$ for the hidden image X can be obtained solving the equation below

$$X^r = A^{-1}Y, \quad (4)$$

The closeness of $X^r$ to X depends on the magnitude of G and properties of matrix A as well. Formally the additive noise can be separated into two components: an exposure independent total noise (see Appendix Eq (A7), later on the name electronic is used as in reports [5,8]) and an exposure dependent conditioned by the fluctuations in the process of the optical photons´ production and detection. For the small pixel mean values when the quantum noise has small contribution in the total pixel noise the shape of the pixel value distribution and the Gaussian are almost alike, that's why later on mainly this form will be used for the electronic noise simulations. The uniform distribution has also been used to show the importance of the distribution form.

The optical photon detection fluctuations depend on the processes of their production, transportation and conversion into an electrical signal. Different from the so-called electronic noise, the exposure dependent noise is increasing with the increase of X-ray photon numbers.

The magnitude of X-ray photon detection fluctuations can be estimated using the Swank factor [9] (here the available reported experimental data (0.95) for 150 µ CsI layer [10] and for the studied device [11] have been used). Performing simple calculations it is possible to show that the relative fluctuation (standard deviation over the mean value (SDM)) in detecting $N_\gamma$ X-ray photons depends on the Swank factor $I_L$ (see Appendix Eq (A6))

$$SDM = \sqrt{\frac{1}{N_\gamma}\left(\frac{1}{I_L} - 1\right)} = \sqrt{\frac{k}{N_p}\left(\frac{1}{I_L} - 1\right)}, \quad (5)$$

where $N_p$ is the pixel raw data value and k is the normalization factor ($N_p = k\,N_\gamma$). Parameter k can be estimated using pixel variance dependence on the pixel mean value [8]. The coefficient of the linear member of the polynomial expansion used in [8] (see Appendix Eq (A7)) depends on the parameter k as well as X-ray photon detection fluctuations. Using simple calculations it can be shown that parameter k can be estimated by the multiplication of the linear member coefficient by the Swank factor (see Appendix Eq (A9)). Here the above-mentioned coefficient value 0.145 from [5] is used and parameter k is estimated (k=0.145x$I_L$).

Considering the case of pixel raw data value

500, the SDM value is about $\cong 0.0037$ (electronic noise value is about $\cong 0.007$ for the same pixel value). Increasing an order of pixel value, the above mentioned values are modified to $\cong 0.0011$ and $\cong 0.0007$ correspondingly. So the exposure dependent fluctuations magnitude for the studied detector is expected to be the same order as the non-dependent one. Thus, later on electronic and exposure dependent noise properties are used for the simulation of G. For the exposure dependent part and for the large pixel values the Gaussian form approximation is good enough. The influence of the additive noise in the restored images has been studied by the simulated images. The given image X is degraded by the matrix A (Eq (2)), then the random noise is added to the resulting image (Eq (3)), and the trial $X^r$ is obtained after the restoration (Eq (4)). The influence of the noise in restoring images is estimated by the distributions of difference ($X^r$-X) and the relative difference ($X^r$-X)/X. The standard deviation of the first distribution is considered as the noise estimation in the restored image. For the real images, X is unknown, so in this case just the noises of Y and $X^r$ images are compared to estimate the noise modification during the restoration.

The influence of $a_{ij}$ errors on the image restoration accuracy is also estimated by the simulations: the image is degraded and then restored using two sets of $a_{ij}$ values. For the degradation of the image, $a_{ij}$ has been generated by Gaussian distributions with mean values and estimation errors (as sigmas) see Table [5] and for the restoration, the mean values from the table are used.

The spatial resolution improvement is studied using the real phantom images acquired by the GE mammography device. The phantom has an absorber edge and LSF is estimated by the edge spread function [12]. The restored image should have resolution conditioned by the physical size of the pixel. Alternatively, the restoration procedure is verified by the estimation of the correlations between the neighbor pixels [5]. The elimination of the above mentioned correlations is considered as an appropriate realization of the restoration procedure.

## III. RESULTS

A series of simulations are carried out for a large range of different noise and pixel values to study the contribution of the additive Gaussian noise in the precision of image restoration. For the image vector X (see Eq (3)) a real breast image of area 160x160 pixels and simulated flat image of the same size are used. The consideration of two different types of images is to show the influence of the image type on the precision of image restoration. Random Gaussian noise is added to the degraded breast image obtained by the application of the matrix A. Here three different varieties of the additive noise are considered. The first one has a standard deviation magnitude similar to the additive electronic noise of the GE mammography device detector [5] (noted as "G_Elec"). The second one (noted as "G_(Exp+Elec)") is the sum of the first one and the exposure dependent noise (SDM is estimated by Eq (5)). The third one has an order larger value than the first one (noted as "large G_Elec"). The resulting distributions of relative differences mentioned in Section 2 for the initial breast image and for different noise values are presented in Fig 2. Note that the SDM (see Section 2) has the value 0.00426 using "G_Elec" noise and 0.00581 when "G_(Exp+Elec)" noise is used. The increase is a little less than $\sqrt{2}$ which means that the exposure dependent part has almost the same contribution in the restoration precision as the exposure independent one (the average pixel value in the breast image is about 885 and has a maximal spread value 552). Taking into account that the use of the Gaussian form for the electronic noise is an approximation, the results of the restoration using the uniformly distributed noise (noted as "U_Elec" with the same variance as "G_Elec") are also shown in Fig 2. As can be seen from the figure, the exposure independent noise distribution shape is less significant (SDM=0.00425).

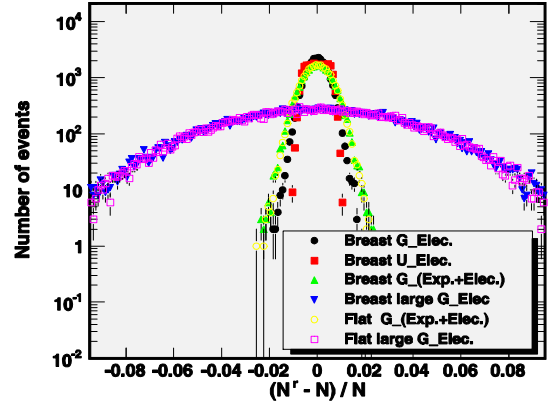

Fig 2 The distributions of the relative differences for the flat and breast initial images.

The same figure also shows the distributions of relative differences for the initial flat image (with pixel value 885) when "G_(Exp+Elec)" and "large G_Elec" noises are used. From Fig 2 it can be appreciated that the flat image restoration precision is similar to non flat one (SDM=0.00573±0.000026). One can also notice from the figure that the large additive noise significantly worsens the image restoration precision. The mean values of these distributions are close to zero (0.00005±$10^{-4}$), which means that the additive noise does not introduce an offset. The restoration precision for the pixel average value (885) is better than

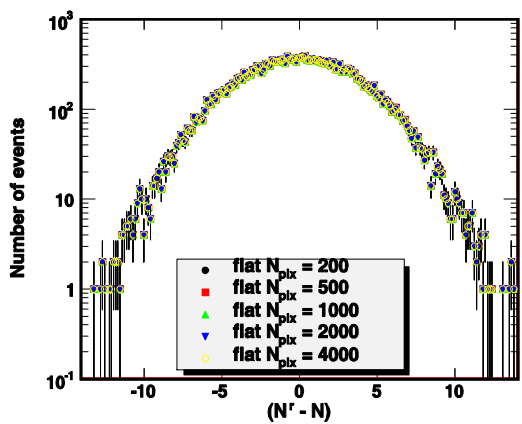

Fig 3 The distributions of differences for the flat images for the different pixel mean values and for the "G_Elec" noise.

±3%. For the larger pixel values the restoration precision should be improved due to Eq (5). For the smaller pixel values the restoration precision worsens because of Eq (5) and the increase of the electronic noise contribution. The influence of the exposure independent noise on the image restoration precision is better studied using simulated flat initial images. The resulting distributions of differences (see Section 2) for the initial flat images with different pixel values and for "G_Elec" noise are presented in Fig 3.

In the restored flat images, the exposure independent additive noise is present in almost the same magnitude as in the original image for the large interval of pixel values (see Fig 3). As to be expected, when the Swank factor approaches unity the absolute precision of restoration is independent of the pixel value. It is also possible to estimate the restoration precision for pixel values smaller than 500 (see Fig 3) where the contribution of exposure dependent noise is less important (see the estimations of SDM in Section 2). Using the results of Fig 3 it can be shown that, better than 3% restoration precision can be achieved for the pixel values larger than 300.

The simulation results show that the restoration is independent of the initial image type and is exact for a smaller area (without noise consideration) within a window which is smaller than the restored area by 12 pixels for each dimension. And ~0.2% accuracy can be reached in the window that is smaller than the restored area by 4 pixels for each dimension.

The influence of $a_{ij}$ estimation errors on the image restoration precision as mentioned in Section 2 has been also estimated by means of simulations. For this purpose, the same breast image as in Fig 2 is used for the vector X (see Eq (3)). The image is degraded by a set of $a_{ij}$ (this set is generated using Gaussian forms see Section 2) and the resulting image is restored by another set of $a_{ij}$ (using mean values shown in the table). The distribution of the relative differences is constructed after repeating the above mentioned procedure 1000 times. These distributions for different $\sigma a_{ij}$ values are shown in Fig 4. As can be seen from the figure, the restoration is sufficiently precise (< 2%) up to $\sigma a_{ij}$ values 0.003 (SDM=0.0061). The accuracy determination of $a_{ij}$ can be done as small as 0.001 [5], which makes it possible to perform the restoration more precisely (<0.5%, SDM = 0.0020). As to be expected these values of SDM are independent of the pixel value. Therefore, the total SDM (including also the additive noise contribution) dependence on the pixel value is conditioned by the additive noise. The same figure also shows that when the "G_(Exp+Elec)" noise is used, the accuracy of the restoration is better than 3% (SDM=0.00605). Comparing this value with the similar one from Fig 2, one can assume that the estimation precision of $a_{ij}$ 0.001 becomes less significant. For the larger pixel values ($N_{pix} \cong 5000$), the $a_{ij}$ estimation errors become as important as the additive noise. So the statement of the less significance of the $a_{ij}$ estimation errors is correct only for the pixel values smaller than 5000. The mean values of distributions in Fig 4 are close to zero (< $0.0003 \pm 10^{-6}$), which means that the estimation errors of $a_{ij}$ do not introduce an offset.

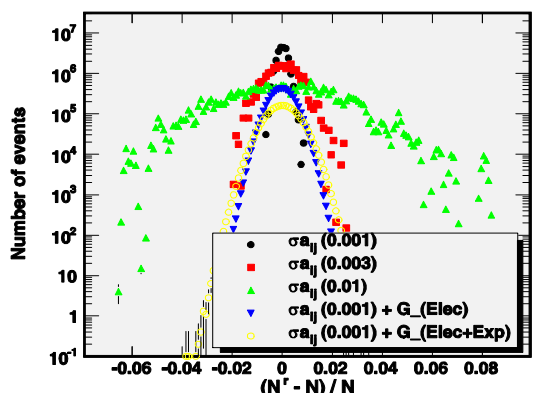

Fig 4 The distributions of the relative differences for the breast initial image without and with additive noise for different values of PSDP estimation errors.

The restoration of an area 16x16 mm of real mammography images is performed to analyze the improvement of an image quality (noise modification, spatial resolution and contrast improvement). The size of the above mentioned image area is constrained by the memory limitations of the computer.

For the noise modification study, flat phantom images for the two different pixel mean values are used. The distributions ($X^r - X^r_{mean}$) and ($Y - Y_{mean}$) (see Section 2) for the pixel mean values 390 and 3660 are shown in Fig 5. The standard deviations of these distributions (see Fig 5) after restoration are 5-7% smaller than the acquired image. For the more precise estimation, it is preferable to use the pair of phantom images acquired in the same conditions to suppress the phantom structure noise [8, 5]. Using pair images for the noise estimation shows that the structure noise in the flat phantom image is negligibly small and the obtained noise modification after restoration is similar to the previous case.

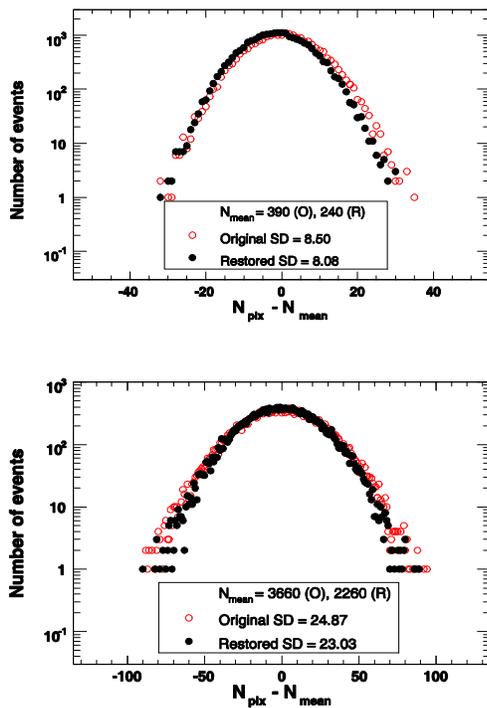

Fig 5 The spread of pixel value of flat phantom images before (O) and after trial restoration (R) for two pixel mean values. The pixel mean values ($N_{mean}$) and the standard deviations (noted as SD) of the distributions are shown inset.

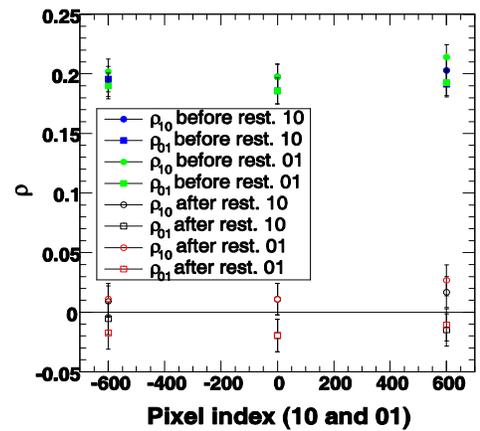

Fig 6 Pixel correlations coefficients in original and restored images measured in different locations and for two different axes (see Fig 1). Zero corresponds to the center of the pixel matrix.

The spatial resolution improvement is studied in two ways (indirect and direct). The decrease of the correlations between neighboring pixels is considered as an indirect way. In the restored flat phantom image, the pixels correlations are eliminated (see Fig 6). The observed large

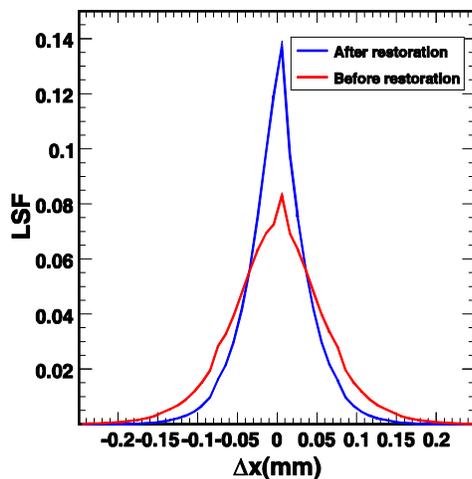

Fig 7 The normalized line spread functions of the statistical errors are due to the used small pixels area [5]. As a direct way for the estimation of spatial resolution improvement, an absorber edge detection method [12] is used. The available flat phantom (containing two plastics each of them having 2 cm thickness and a rhodium foil in the midst) has been constructed to estimate the foil thickness measurement. The averaged LSF that is obtained by the differentiation of edge spread function along the image rows before and after restoration is shown in Fig 7. The standard deviation of LSF for the restored image is of 0.034 mm (the expected value for the ideal detector with pixel size 0.1 mm is 0.029 mm). This difference can be explained by the following contributions: the spread from the small scatterings in plastics (because of the Greed); the spread from the focal spot and the spread from the non

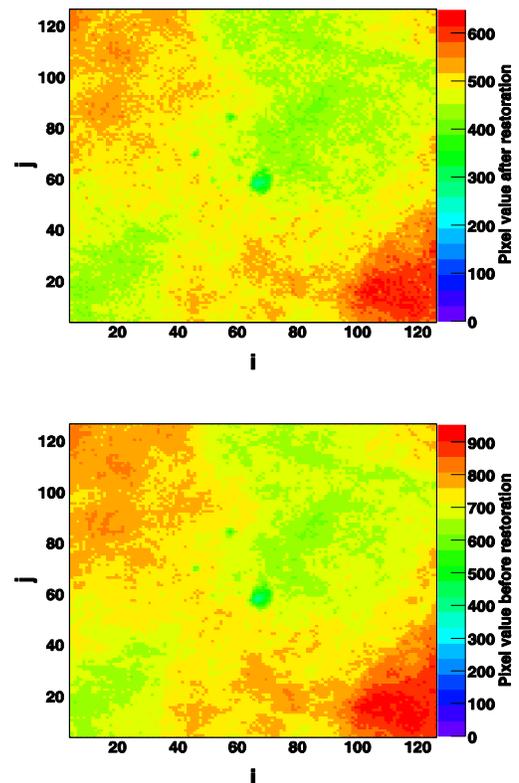

Fig 8 An area of a breast image with microcalcifications: bottom-before and top-after sharpness of the foil edge.

The restored breast image (having

microcalcifications) with the original one is shown in Fig 8. The contrast improvement for the microcalcifications (is within 20-40%) agrees with the expectation for the small pixel size objects.

The restoration time for the image with sizes 160x160 pixels using Gaussian elimination with non time optimize program is about a few seconds for Pentium 2.2 GhZ machine. For the real images having $10^2$ more points, it is necessary to use the fast algorithms [1].

## IV. Discussion

The restoration process depends on the properties of the matrix A (the number of diagonals with non zero elements as well as the magnitude of these elements and their estimation errors). The estimation errors of the matrix elements obtained in the study [5] can be considered acceptable for the precise restoration. The additive noise will not introduce an observable offset in the restored image and its magnitude is only slowly modified.

The obtained result of the restoration precision is better than (3%) for the pixel values larger than 300 (real breast images usually have pixel values larger than 300).

The resolution improvement agrees with the expectation, though more precise measurements of LSF will be better to perform a more precise evaluation. The correlation elimination can be considered as an alternative to LSF measurements which is less sensitive to quantum noise magnitude. This allows an easy computer control of the restoration process. In fact, the whole procedure starting from the pixel simultaneous detection probabilities up to the final restoration can be done in automatic mode. This is important for the image quality control and for the detector design as well.

The possibility of the "exact" (without using minimization procedure) de-blurring of the degradations introduced by the detector somehow can solve the resolution problem in case of using thick converters [6] in the detectors based on the indirect detection method. High efficiency is provided by the thick converters which makes possible the decrease of the dose value for the patient.

For the real time on-line application, it is necessary to use faster methods than the Gaussian elimination or faster computers having the productivity more than one order larger than the used one. Here in this study, the main purpose was to analyze the matrix obtained using pixel simultaneous probabilities as well as to estimate the required magnitude of errors for the appropriate restoration.

V. CONCLUSIONS

The possibility of the "exact" restoration of the detector degradation can solve the problem of the thick converters usage. The use of pixel simultaneous detection probabilities for this purpose is suggested to be a more convenient choice.

The additive noise is present in the restored images and almost has the same magnitude as in the original. The estimation errors of pixel simultaneous detection probabilities (for the GE Senographe 2000D device detector) allow restoring images with the accuracy better than 1%. Introducing the additive noise, the above mentioned precision worsens up to (3%). The restoration (without noise consideration) of the local area is exact in the window which is smaller than the restored one (12 pixels) for each dimension. In the restored small area (16x16 mm) of real images, the pixel responses are not correlated, which can be considered as an alternative independent check-up of the restoration process. The spatial resolution improvement agrees with the expected one. The contrast improvement is 20-40% for the small objects and agrees with the estimations.

ACKNOWLEDGMENT

The author is grateful to the referee for his valuable comments that greatly improved an earlier version of this paper; to M-E Brandan and Y. Villaseñor, for kindly providing access to the mammography unit; to radiological technicians of the National Institute of Cancer Research for technical support; to M. Grabska for preparation of the text; and PAPIT-UNAM IN-115409 for the partial support.

APPENDIX

The Swank factor $I_L$ is defined for a given distribution as [9]:

$$I_L = \frac{m_1^2}{m_0 m_2}, \quad (A1)$$

where $m_0$, $m_1$, and $m_2$ are respectively the zeroth, first, and second moments of the distribution of the random variable N (in our case N is the single X-ray detection signal value). For the normalized distribution $m_0$ will have a unit value and using the definitions of the moments the expression can be written [13].

$$I_L = \frac{E[N]^2}{E[N^2]} = \frac{\mu^2}{V(N)+\mu^2}, \quad (A2)$$

where E[] signifies the mathematical expectation, V(N) is the variance of N and µ is the average value. Taking into account that V(N) is the square of the standard deviation (SD) and using Eq (A2), it can be obtained that

$$SD^2 = \mu^2 \left(\frac{1}{I_L} - 1\right), \quad (A3)$$

Now considering a sum of n variables having the same variances and mean values ($S=N_1+N_2+.....+N_n$) as in case of n X-ray photon detection, the variance of V(S) can be written as

$$V(S) = \sum_i^n V(N_i) = nSD^2, \quad (A4)$$

Combining Eq (A3) and (A4) for the V(S) it can be shown that

$$V(S) = n\mu^2 \left(\frac{1}{I_L} - 1\right), \quad (A5)$$

Now defining the relative variation as standard deviation over the mean value (SDM) and using Eq (A5) it can be obtained that

$$SDM \equiv \frac{\sqrt{V(S)}}{n\mu} = \sqrt{\frac{1}{n}\left(\frac{1}{I_L} - 1\right)}, \quad (A6)$$

So in case of detection of n X-ray photons the relative fluctuations of the sum signal decreases as $\sqrt{1/n}$.

The scaling factor for the photon signal transition can be estimated by means of the pixel variance behavior. The pixel variance dependence on the pixel raw data mean value $N_p$ can be represented [8] as.

$$V(N_p) = a_0 + a_1 N_p + a_2 N_p^2 \quad (A7)$$

Here $a_1$ depends on the scaling factor and the fluctuations of detecting an X-ray photon. Therefore, the linear member of total variance ($a_1 N_p$) is the sum of the quantum and the above-mentioned fluctuation variances.

$$a_1 N_p = V(kN_\gamma) + SDM^2(kN_\gamma)^2 \quad (A8)$$

Where $N_\gamma$ is the photon mean value corresponding to $N_p$, and k is the scaling factor ($N_p=kN_\gamma$), SDM is defined by Eq (A6). Dividing both parts of Eq A8 over $N_p^2$ and using

Eq (A6) and connections ($N_p = kN_\gamma$ and $V(kN_\gamma) = k^2 N_\gamma$) it can be seen that

$$\frac{a_1}{k} = 1 + \left(\frac{1}{I_L} - 1\right) \xrightarrow{yields} k = a_1 I_L. \quad (A9)$$

This relation can be considered as an alternative manner to estimate the Swank factor.

MTF of digital radiographic systems using an edge test device, *Med. Phys.* 25 (1998): 102–113.